\documentclass[conference]{IEEEtran}
\IEEEoverridecommandlockouts
\usepackage{cite}
\usepackage{graphicx}
\usepackage{amsmath}
\usepackage{times}
\usepackage{latexsym}
\usepackage{bm}
\usepackage{amssymb}
\usepackage[center]{caption2}
\usepackage{array}
\usepackage{fancyhdr}
\usepackage{cite,graphicx,amsmath,amssymb}
\usepackage{citesort}
\usepackage{psfrag}
\usepackage{multirow}

\ifCLASSINFOpdf

\else

\fi

\newtheorem{prop}{Proposition}
\newtheorem{alg}{Algorithm}
\newtheorem{Definition}{Definition}

\newenvironment{sequation}{\begin{equation}\small}{\end{equation}}

\newcommand{\tr}{\text{tr}}

\makeatother

\begin{document}
\title{Linear MIMO Precoding in Jointly-Correlated
Fading Multiple Access Channels with Finite Alphabet Signaling}

\author{\IEEEauthorblockN{Yongpeng Wu, Chao-Kai Wen, Chengshan Xiao, Xiqi Gao, and Robert Schober}

\thanks{The work of Y. Wu and R. Schober was supported
by the Alexander von Humboldt Foundation. The work of C.-K. Wen was supported by
the National Science Council, Taiwan, under grant NSC100-2221-E-110-052-MY3.
The work of C. Xiao was supported in part by the National Science Foundation under grant ECCS-1231848.
The work of X. Gao was supported by National Natural Science Foundation of China under Grants 61320106003 and 61222102,
the China High-Tech 863 Plan under Grant 2012AA01A506, National Science and Technology Major Project of China under Grants 2013ZX03003004
and 2014ZX03003006-003, and the Program for Jiangsu Innovation Team.
}

\thanks{Y. Wu and R. Schober are with Institute for Digital Communications, Universit$\ddot{a}$t Erlangen-N$\ddot{u}$rnberg,
Cauerstrasse 7, D-91058 Erlangen, Germany (Email: yongpeng.wu@lnt.de; schober@lnt.de). Y. Wu was with the 
National Mobile Communications Research Laboratory,
Southeast University, Nanjing, P. R. China (Email: ypwu@seu.edu.cn).}

\thanks{C. K. Wen is with the Institute of Communications Engineering, National Sun Yat-sen University, Kaohsiung 804,
Taiwan (Email:  chaokaiwen@gmail.com).}

\thanks{C. Xiao is with the Department of Electrical and Computer Engineering,
Missouri University of Science and Technology, Rolla, MO 65409, USA (Email: xiaoc@mst.edu). }

\thanks{X. Gao is with the National Mobile Communications Research Laboratory,
Southeast University, Nanjing, 210096, P. R. China (Email: xqgao@seu.edu.cn). }

}

\maketitle

\begin{abstract}
In this paper, we investigate the design of linear precoders for 
multiple-input multiple-output (MIMO) multiple access channels (MAC).
We assume that statistical channel state information (CSI) is available
at the transmitters and consider the problem under the practical finite alphabet
input assumption. First, we derive an asymptotic (in the large-system limit)
weighted sum rate (WSR) expression for the MIMO MAC with finite alphabet inputs
and  general jointly-correlated fading. Subsequently, we
obtain necessary conditions for  linear precoders maximizing the
asymptotic WSR and propose
an iterative algorithm for determining the precoders of all users.
In the proposed algorithm, the search space of each user for designing the precoding matrices
is its own modulation set.  This significantly reduces the dimension of the search space
for finding the precoding matrices of all users
compared to the conventional precoding design for the MIMO MAC  with finite alphabet inputs, where the search
space is the combination of the modulation sets of all users. As
a result, the proposed algorithm decreases the computational complexity
for MIMO MAC precoding design with finite alphabet inputs by several orders of magnitude.
Simulation results for finite alphabet signalling  indicate that
the proposed iterative algorithm achieves
significant performance gains over  existing precoder designs, including
the precoder design based on the Gaussian input assumption, in terms of both the sum rate and the coded bit error rate.
\end{abstract}


\section{Introduction}
In recent years, the channel capacity and the design of optimum
transmission strategies for multiple-input multiple-output (MIMO) multiple access channels (MAC)
have been widely studied \cite{goldsmith2003capacity,Yu2002}.
However, most works on MIMO MAC rely on the critical assumption
of Gaussian input signals.  Although Gaussian inputs are optimal in theory, they are rarely used 
in practice. Rather, it is well-known that practical communication signals usually are drawn from finite constellation sets,
such as pulse amplitude modulation (PAM), phase shift keying (PSK) modulation, and quadrature amplitude modulation (QAM). These finite constellation sets
differ significantly from the Gaussian idealization \cite{Lozano2006TIT}.  Accordingly,  transmission schemes designed based on the Gaussian input assumption may
result in substantial performance losses when finite alphabet inputs are used for transmission\cite{Xiao2011TSP,Wang2011,Wu2012TWC,Wu2012TVT,Wu2013TCOM}.
For the case of the two-user single-input single-output MAC with finite alphabet inputs, the optimal angle of rotation and the optimal power division
between the transmit signals were found in \cite{Harshan2010TIT} and \cite{harshan2013novel}, respectively.
For the MIMO MAC with an arbitrary number
of users and generic antenna configurations,
 an iterative algorithm for searching for the optimal precoding matrices of all users was proposed in \cite{Wang2011}.

The transmission schemes in \cite{Xiao2011TSP,Wang2011,Wu2012TWC,Wu2012TVT,Wu2013TCOM,Harshan2010TIT,harshan2013novel,Wang2011}
require accurate instantaneous channel state information (CSI) available
at the transmitters for precoder design. However, in some applications,
the obtained instantaneous CSI at the transmitters might be outdated. Therefore, for these scenarios, it is more
reasonable to exploit the channel statistics at the transmitter for precoder design,
as they change much slower
than the instantaneous channel parameters \cite{wen2011sum}.
For finite alphabet inputs, for point-to-point systems,
an efficient precoding algorithm for maximization of the ergodic capacity over Kronecker fading channels was developed in \cite{zeng2012linear}.
Also, in \cite{wen2007asymptotic}, asymptotic (in the large-system limit) expressions for the mutual information
of the MIMO MAC with Kronecker fading were derived. Despite these previous works, the study of the MIMO MAC
with statistical CSI at the transmitter and finite alphabet inputs remains incomplete, for two reasons: First, the
Kronecker fading model characterizes the correlations of the transmit and the receive antennas separately, which is often not in agreement with
measurements \cite{weichselberger2006stochastic}.  In contrast,  jointly-correlated fading models, such as
 Weichselberger's model \cite{weichselberger2006stochastic}, do not only account for the correlations at both ends of the link, but
also characterize their mutual dependence. As a consequence,  Weichselberger's model
provides a more general representation of MIMO channels. Second, systematic precoder designs for statistical CSI at the transmitter
for the MIMO MAC with finite alphabet inputs have not been reported yet, even for the Kronecker fading model.

In this paper, we investigate the linear precoder design for the $K$-user MIMO MAC assuming Weichselberger's fading model,
finite alphabet inputs, and availability of statistical CSI at the transmitter.
By exploiting a random matrix theory tool from statistical physics, called the replica
method\footnote{We note that the replica method has been applied to communications problems
before \cite{wen2007asymptotic,Tanaka2002TIT,RMuller2008JSAC}.},
we first derive an asymptotic expression for
the weighted sum rate (WSR) of the MIMO MAC for Weichselberger's fading model in the large-system regime 
where the numbers of transmit and receive antenna both approach infinity.
The derived expression indicates that the WSR
can be obtained asymptotically by calculating the mutual information of each user separately
over equivalent deterministic channels.
This property significantly reduces the computational effort for calculation of the WSR.
Furthermore, exploiting Karush-Kuhn-Tucker (KKT) analysis,
we establish necessary conditions for the optimal precoding
matrices for asymptotic WSR maximization.
This analysis facilities the derivation of an efficient iterative gradient descent
algorithm\footnote{It is noted that although we derive the asymptotic WSR in the large-system regime,
the proposed algorithm can also be applied for systems with a finite number of antennas.} for finding the optimal precoders of all users.
In the proposed algorithm,
the search space for the design of the precoding matrix of each user is only the user's own modulation set.
Accordingly, denoting the number of transmit antennas
and the size of the modulation set of user $k$ by $N_t$ and
$Q_k$, respectively,  the dimensionality of the search space for finding the precoding matrices of all users
with the proposed algorithm
is $\sum\nolimits_{k = 1}^K {Q_k^{2 N_t } }$,  whereas the dimensionality of the search space
of the algorithm employing instantaneous CSI at the transmitter in \cite{Wang2011}
is $\left( {\prod\nolimits_{k = 1}^K {Q_k } } \right)^{ 2 N_t }$.
This indicates that the proposed algorithm does not only provide a systematic precoder design method for the MIMO MAC
with  statistical CSI at the transmitter, but also reduces the implementation complexity
by \emph{several orders of magnitude} compared to the precoder design for instantaneous CSI.
Moreover, the precoder designed for statistical CSI
has to be updated much less frequently than the precoder designed for instantaneous CSI as the channel statistics change very slowly compared
to the instantaneous CSI.
In addition, unlike the algorithm in \cite{Wang2011}, the proposed algorithm does not require the computationally expensive
simulation over each channel realization.
Numerical results demonstrate that the proposed design  provides substantial performance gains over
systems without precoding and systems employing precoders designed under the Gaussian input assumption.

The following notations are adopted throughout the paper:  Column vectors are represented by lower-case bold-face letters,
and matrices are represented by upper-case bold-face letters. Superscripts $(\cdot)^{T}$, $(\cdot)^{*}$, and $(\cdot)^{H}$
stand for the matrix/vector transpose, conjugate, and conjugate-transpose operations, respectively.
$\rm{det}(\cdot)$ and $\rm{tr}(\cdot)$  denote the matrix determinant and trace operations, respectively. ${\rm{diag}}\left\{\bf{b}\right\}$ and ${\rm{blockdiag}}\left\{{\bf{A}}_k\right\}_{k=1}^{K}$   denote
diagonal matrix and block diagonal matrix containing
in the main diagonal (or block diagonal) the elements of vector $\bf{b}$ and matrices ${\bf{A}}_k, k=1,2,\cdots,K$, respectively. $\odot$ and $\bigotimes$ denote the element-wise product
and the Kronecker product of two matrices. ${\rm vec} \left(\mathbf{A}\right) $ returns a column vector whose
entries are the ordered stack of columns of $\mathbf{A}$. $[\mathbf{A}]_{mn}$ denotes the element in the
$m$th row and $n$th column of matrix $\mathbf{A}$.
$\left\| {\mathbf{X}} \right\|_F$ denotes the Frobenius norm of matrix $\mathbf{X}$.
${\mathbf{I}}_M$ denotes an $M \times M$ identity matrix,
and $E_V\left[\cdot \right]$ represents the expectation with respect to random variable $V$, which can be a scalar, vector, or matrix.

\section{System Model}\label{sec:model}
Consider a MIMO MAC system with $K$ independent users. We suppose each of the $K$
users has $N_t$ transmit antennas and the receiver has $N_r$ antennas.  Then, the received signal
$\mathbf{y} \in \mathbb{C}^{N_r \times 1}$ is given by
\begin{equation}\label{model}
{\bf{y}} = \sum\limits_{k = 1}^K {{\bf{H}}_k {\bf{x}}_k }  + {\bf{v}}
\end{equation}%
where ${\mathbf{x}_k} \in \mathbb{C}^{N_t \times 1}$ and ${\mathbf{H}}_k \in \mathbb{C}^{N_r \times N_t}$ denote the transmitted signal
and the channel  matrix of user $k$, respectively.
${\mathbf{v}} \in \mathbb{C}^{N_r \times 1}$ is a zero-mean complex Gaussian noise vector
with covariance matrix\footnote{To simplify our notation, in this paper, without loss of generality, we
normalize the power of the noise to unity.} $\mathbf{I}_{N_r}$.
Furthermore, we make the common assumption (as e.g. \cite{Soysal2007JSAC,wen2011sum})
that the receiver has the instantaneous CSI of all users, and each transmitter has the statistical CSI of all users.

The transmitted signal vector ${\mathbf{x}_k}$ can be expressed as
\begin{equation}\label{x_signal}
{\mathbf{x}_k} = {{\mathbf{B}}_k {\mathbf{d}}_k }
\end{equation}%
where ${\mathbf{B}}_k$ and ${\mathbf{d}}_k$ denote the linear precoding matrix and the
input data vector of user $k$, respectively. Furthermore, we assume ${\mathbf{d}}_k$ is a zero-mean
vector with covariance matrix $\mathbf{I}_{N_t}$. Instead of employing the traditional assumption
of a Gaussian transmit signal, here we assume ${\mathbf{d}}_k$ is taken from
a discrete constellation, where all elements of the constellation are equally likely.
In addition, the transmit signal ${\mathbf{x}_k}$
conforms to the power constraint
\begin{equation}\label{x_constraint_2}
 E_{\mathbf{x}_k}\left[ {\mathbf{x}_k^H {\mathbf{x}_k}} \right] = {{\rm{tr}}\left( {{\mathbf{B}}_k {\mathbf{B}}_k^H} \right)} \leq P_k, \ k = 1,2,\cdots,K.
\end{equation}%

For the jointly-correlated fading MIMO channel, we adopt Weichselberger's model \cite{weichselberger2006stochastic}
throughout this paper. This model jointly characterizes the correlation
at the transmitter and receiver side \cite{weichselberger2006stochastic}. In particular, for user $k$,
$\mathbf{H}_k$ can be modeled as \cite{weichselberger2006stochastic}
\begin{equation}\label{H_channel}
{\bf{H}}_k  = {\bf{U}}_{{\rm{R}}_k } \left( {{\bf{ \widetilde{G}}}_k  \odot {\bf{W}}_k } \right){\bf{U}}_{{\rm{T}}_k }^H
\end{equation}%
where ${\bf{U}}_{{\rm{R}}_k }  = \left[ {{\bf{u}}_{{\rm{R}}_k ,1} ,{\bf{u}}_{{\rm{R}}_k ,2} , \cdots ,{\bf{u}}_{{\rm{R}}_k ,N_r} } \right] \in \mathbb{C}^{N_r \times N_r}  $ and ${\bf{U}}_{{\rm{T}}_k }  = \left[ {{\bf{u}}_{{\rm{T}}_k ,1} ,{\bf{u}}_{{\rm{T}}_k ,2} , \cdots ,{\bf{u}}_{{\rm{T}}_k ,N_t} } \right] \in \mathbb{C}^{N_t \times N_t}$ represent
deterministic unitary matrices, respectively.  ${\bf{ \widetilde{G}}}_k \in \mathbb{C}^{N_r \times N_t} $ is a  deterministic matrix with real-valued nonnegative elements,
and ${\bf{W}}_k \in \mathbb{C}^{N_r \times N_t}$ is a random matrix with independent identically distributed (i.i.d.) Gaussian elements with zero-mean and unit variance.
We define  ${\bf{G}}_k  = {\bf{\widetilde{{G}}}}_k  \odot {\bf{ \widetilde{{G}}}}_k$ and let $g_{k,n,m}$ denote the $(n,m)$th element of matrix
${\bf{G}}_k$. Here $\mathbf{G}_k$ is referred to ``coupling matrix"  as $g_{k,n,m}$ corresponds to the average coupling energy
between ${\bf{u}}_{{\rm{R}}_k ,n}$ and ${\bf{u}}_{{\rm{T}}_k ,m}$\cite{weichselberger2006stochastic}.  Henceforth, the transmit and
receive correlation matrices of user $k$ can be written as
\begin{equation}\label{correlation}
\begin{array}{l}
 {\bf{R}}_{t,k}  = E_{{\bf{H}}_k}\left[ {{\bf{H}}_k^H {\bf{H}}_k } \right] = {\bf{U}}_{{\rm{T}}_k } \boldsymbol{\Gamma} _{{\rm{T}}_k } {\bf{U}}_{{\rm{T}}_k }^H  \\
 {\bf{R}}_{r,k}  = E_{{\bf{H}}_k}\left[ {{\bf{H}}_k {\bf{H}}_k^H } \right] = {\bf{U}}_{{\rm{R}}_k } \boldsymbol{\Gamma} _{{\rm{R}}_k } {\bf{U}}_{{\rm{R}}_k }^H  \\
 \end{array}
\end{equation}
where  $\boldsymbol{\Gamma}_{{\rm{T}}_k } $ and  $\boldsymbol{\Gamma}_{{\rm{R}}_k }$ are diagonal matrices with $
\left[ {{\bf{\Gamma }}_{{\rm{T}}_k } } \right]_{mm}  = \sum\nolimits_{n = 1}^{N_r } {g_{k,n,m} }$, $m = 1,2,\cdots,N_t$ and $
\left[ {{\bf{\Gamma }}_{{\rm{R}}_k } } \right]_{nn}  = \sum\nolimits_{m = 1}^{N_t } {g_{k,n,m} }$, $n = 1,2,\cdots,N_r$, respectively.

\section{Asymptotic WSR of MIMO MAC with Finite Alphabet Inputs}
We divide all users into two groups, denoted as set $\cal A$ and its
complement set ${\cal{A}}^c$: ${\cal {A}} =\{i_1, i_2, \cdots, i_{K_1} \}\subseteq \{1,2, \cdots, K\}$ and ${\cal {A}}^c = \{j_1, j_2, \cdots, j_{K_2}\}$,
$K_1+K_2= K$. Also, we define ${\bf{H}}_{\cal {A}} = \left[ {{\bf{H}}_{i_1} \ {\bf{H}}_{i_2}  \cdots \mathbf{H}_{i_{K_1}} } \right]$,
$\mathbf{d}_{\cal {A}} = \left[ \mathbf{d}_{i_1}^T \ \mathbf{d}_{i_2}^T  \cdots \mathbf{d}_{i_{K_1}}^T \right]^T$,
$ \mathbf{d}_{{\cal {A}}^c} = \left[ \mathbf{d}_{j_1}^T \ \mathbf{d}_{j_2}^T  \cdots \mathbf{d}_{j_{K_2}}^T \right]^T$,
$\mathbf{B}_{\cal {A}} = {\rm{blockdiag}} \left\{ \mathbf{B}_{i_1},\mathbf{B}_{i_2}, \cdots, \mathbf{B}_{i_{K_1}} \right\}$,
and ${\bf{y}}_{\cal {A}} = {\bf{H}}_{\cal {A}} \mathbf{B}_{\cal {A}} \mathbf{d}_{\cal {A}} + {\bf{v}}$. Then, the capacity region $(R_1, R_2, \cdots, R_K)$ of the $K$-user MIMO MAC satisfies the following
conditions\cite{Cover}:
\begin{sequation}\label{capacity_region}
    \sum_{i\in {\cal {A}}} R_i \leq I
    \left( \mathbf{d}_{\cal {A}}; \mathbf{y} | \mathbf{d}_{{\cal {A}}^c} \right), \quad \forall {\cal {A}}\subseteq \{ 1, 2, \cdots, K \}
\end{sequation}%
where
\begin{sequation}\label{mutual_info_1}
I\left( {{\bf{d}}_{\cal {A}} ;{\bf{y}}\left| {{\bf{d}}_{{\cal A}^c } } \right.} \right) \!\! = \!\! E_{{\bf{H}}_{\cal {A}} } \left[ {E_{{\bf{d}}_{\cal {A}} ,{\bf{y}}_{\cal {A}} } \left[ {\log _2 \frac{{p\left( {{\bf{y}}_{\cal {A}} \left| {{\bf{d}}_{\cal {A}} {\bf{,H}}_{\cal {A}} } \right.} \right)}}{{p\left( {{\bf{y}}_{\cal {A}} \left| {{\bf{H}}_{\cal {A}} } \right.} \right)}}\left| {\bf{H}} \right._{\cal {A}} } \right]} \right].
\end{sequation}%
In (\ref{mutual_info_1}), ${p} ({\bf{y}}_{\cal {A}} |{\bf{H}}_{\cal {A}})$ denotes the marginal probability density function (p.d.f.) of $p(\mathbf{d}_{\cal {A}},\mathbf{y}_{\cal {A}}|\mathbf{H}_{\cal {A}})$. As a result, we have
\begin{sequation}\label{eq:Finite_Mutual}
\begin{array}{l}
I (\mathbf{d}_{\cal {A}};\mathbf{y} \left| {{\bf{d}}_{A^c } } \right. ) = \\
 \hspace{-0.2cm} - E_{\mathbf{H}_{\cal {A}}}\!\left[ \! E_{\mathbf{y}_{\cal {A}}} \!\left[\!\left.\log_2 E_{\mathbf{d}_{\cal {A}}}\left[e^{-\left\|\mathbf{y}_{\cal {A}}-\mathbf{H}_{\cal {A}}  \mathbf{B}_{\cal {A}} \mathbf{d}_{\cal {A}}\right\|^2}\right]\right|\mathbf{H}_{\cal {A}}\! \right] \! \right] \!-\! N_r \log_2 e.
\end{array}
\end{sequation}%
The expectation in (\ref{eq:Finite_Mutual}) can be evaluated numerically by Monte-Carlo simulation. However,
for a large number of antennas, the computational complexity could be enormous. Therefore,
by employing the replica method, a classical technique from statistical physics, we obtain an asymptotic expression
for (\ref{eq:Finite_Mutual}) as detailed in the following.

\subsection{Some Useful Definitions}
We first provide some useful definitions. Consider a virtual MIMO channel defined by
\begin{sequation}\label{eq:EqScalGAUEach}
\mathbf{z}_{\cal {A}}= \sqrt{\mathbf{T}_{\cal {A}}} \mathbf{B}_{\cal {A}} \mathbf{d}_{\cal {A}} + \check{\bf v}_{\cal {A}}
\end{sequation}%
$\mathbf{T}_{\cal A} \in \mathbb{C}^{K_1 N_t \times K_1 N_t} $ is
given by $\mathbf{T}_{\cal A} =  {\rm{blockdiag}}$ $\left( \mathbf{T}_{i_1}, \mathbf{T}_{i_2}, \dots, \mathbf{T}_{i_{K_1}} \right)
\in\mathbb{C}^{K_1 N_t \times K_1 N_t}$, where $\mathbf{T}_{i_k}\in\mathbb{C}^{N_t \times N_t}$ is a
deterministic matrix, $k = 1,2,\cdots,K_1$. $\check{\bf v}_{\cal A} \in \mathbb{C}^{K_1 N_r \times 1} $  is a standard complex Gaussian random vector with i.i.d. elements. The minimum mean square error (MMSE) estimate of  signal vector $\mathbf{d}_{\cal A}$ given (\ref{eq:EqScalGAUEach}) can be expressed
as
\begin{sequation}\label{eq:hatx_k}
 \hat{\mathbf{d}}_{\cal {A}} = E_{\mathbf{d}_{\cal {A}}} \left[ E_{\check{\bf v}_{\cal {A}}}\left[\mathbf{d}_{\cal {A}} \Big|\mathbf{z}_{\cal {A}},\sqrt{\mathbf{T}_{\cal {A}}},\mathbf{B}_{\cal {A}} \right] \right].
 \end{sequation}%
 Define the following mean square error (MSE) matrix
 \begin{sequation}\label{eq:mse}
    \boldsymbol{\Omega}_{\cal {A}} = \mathbf{B}_{\cal {A}} E_{\mathbf{z}_{\cal {A}}} \left[ E_{\mathbf{d}_{\cal {A}}}\left[( \mathbf{d}_{\cal {A}} - \hat{\mathbf{d}}_{\cal {A}}) (\mathbf{d}_{\cal {A}} - \hat{\mathbf{d}}_{\cal {A}})^H \right] \right] \mathbf{B}_{\cal {A}}^H.
\end{sequation}%
Also, define the MSE matrix of the $i_k$th ($i_1 \leq i_k \leq i_{K_1}$)
user as
\begin{equation*}
  \boldsymbol{\Omega}_{i_k} =  \langle{\boldsymbol{\Omega}}_{\cal A}\rangle_k
\end{equation*}
where $\langle \mathbf{X} \rangle_k \in \mathbb{C}^{N_t \times N_t} $ denotes a submatrix obtained by extracting
the $\left( (k - 1) N_t  + 1 \right)$th to the $(k N_t)$th row and column elements of matrix $\mathbf{X}$.

\begin{Definition}\label{Definition_1}
Define vectors $\boldsymbol{\gamma}_{i_k} = [\gamma_{i_k,1}, \gamma_{i_k,2} ,\dots, \gamma_{i_k,N_r}]^T$ and $\boldsymbol{\psi}_{i_k} = [\psi_{i_k,1}, \psi_{i_k,2}
,\dots, \psi_{i_k,N_t}]^T$. Define the following matrices
\begin{sequation} \label{eq:eqChMatrixTR}
\left\{\begin{aligned}
\mathbf{T}_{i_k} & =\mathbf{U}_{{\rm T}_{i_k}}{\rm diag}\left( \mathbf{G}_{i_k}^T \boldsymbol{\gamma}_{i_k} \right)\mathbf{U}_{{\rm T}_{i_k}}^{H}\in\mathbb{C}^{N_t \times N_t}\\
\mathbf{R}_{i_k} &= \mathbf{U}_{{\rm R}_{i_k}}{\rm diag}\left( \mathbf{G}_{i_k} \boldsymbol{\psi}_{i_k}\right)\mathbf{U}_{{\rm R}_{i_k}}^{H}\in\mathbb{C}^{N_r \times N_r}
\end{aligned}\right.
\end{sequation}%
where $\gamma_{i_k,n}$ and $\psi_{i_k,m}$ satisfy the following equations
\begin{sequation}\label{eq:Varsigma_k-MSE}
\left\{\begin{aligned}
\gamma_{i_k,n} &=  \mathbf{u}_{{\rm R}_{i_k},n}^{H} \left(\mathbf{I}_{N_r}+\mathbf{R}_{\cal {A}}\right)^{-1} \mathbf{u}_{{\rm R}_{i_k},n}, \ n = 1,2,\cdots,N_r \\
\psi_{i_k,m}   &= \mathbf{u}_{{\rm T}_{i_k},m}^{H} \boldsymbol{\Omega}_{i_k} \mathbf{u}_{{\rm T}_{i_k},m}, \  m = 1,2,\cdots,N_t
\end{aligned}
\right.
\end{sequation}%
\end{Definition}

\subsection{Asymptotic Mutual Information}
Now, we are ready to provide a simplified asymptotic expression of (\ref{eq:Finite_Mutual}).
\begin{prop}\label{prop:ach_rate_mac}
 For the MIMO MAC model (\ref{model}), when $N_r$ and $N_t$ both approach
infinity but the ratio
$\beta = N_t/ N_r$ is fixed, the mutual information in (\ref{eq:Finite_Mutual}) can be asymptotically
 approximated\footnote{It it noted that the asymptotic expression obtained based on the replica method is also useful for
 systems with a finite number of antennas \cite{wen2007asymptotic}.} by
\begin{sequation}\label{eq:GAUMutuall}
\begin{array}{l}
I\left( {{\bf{d}}_{\cal {A}} ;{\bf{y}}\left| {{\bf{d}}_{ {\cal {A}}^c } } \right.} \right) \simeq  I\left( \mathbf{d}_{\cal {A}};\mathbf{z}_{\cal {A}} \big| {{ \sqrt{\mathbf{T}_{\cal {A}}} \mathbf{B}_{\cal {A}}}} \right) \\
\hspace{0.5cm}  + \log_2 \det\left(\mathbf{I}_{N_r}+\mathbf{R}_{\cal {A}}\right) - \log_2 e \sum_{k=1}^{K_1} \boldsymbol{\gamma}_{i_k}^T \mathbf{G}_{i_k} \boldsymbol{\psi}_{i_k}
\end{array}
\end{sequation}%
where $I\left( \mathbf{d}_{\cal A};\mathbf{z}_{\cal A} \big| {{ \sqrt{\mathbf{T}_{\cal A}} \mathbf{B}_{\cal A}}} \right)$ represents the mutual information
between $\mathbf{d}_{\cal A}$ and $\mathbf{z}_{\cal A}$ of channel model (\ref{eq:EqScalGAUEach}).
\begin{proof}
Please refer to Appendix \ref{sec:proof_ach_rate_mac}.
\end{proof}
\end{prop}

Suppose the transmit signal ${\mathbf{d}}_k$ is taken from a discrete constellation with cardinality
$Q_k$. Define $M_k = {Q_k}^{N_t}$. $S_k$ denotes the constellation set of user $k$.
 ${\mathbf{a}}_{k,j}$ denotes the $j$th element in the constellation set $S_k$, $k =1,2,\cdots ,K$,
 $j = 1,2, \cdots,M_k$. Then, based on the definition of $\mathbf{T}_{\cal {A}}$, $\mathbf{B}_{\cal {A}}$, and model
(\ref{eq:EqScalGAUEach}), (\ref{eq:GAUMutuall}) can be further simplified as
\begin{sequation}\label{eq:GAUMutuall_2}
\begin{array}{l}
I\left( {{\bf{d}}_{\cal {A}} ;{\bf{y}}\left| {{\bf{d}}_{{\cal {A}}^c  } } \right.} \right) \simeq
\sum\limits_{i \in {\cal {A}}} {I\left( {{\bf{d}}_{i_k } ;{\bf{z}}_{i_k } \left| {\sqrt {{\bf{T}}_{i_k } } {\bf{B}}_{i_k } } \right.} \right)} \\
\hspace{0.5cm} + \log_2 \det\left(\mathbf{I}_{N_r}+\mathbf{R}_{\cal {A}}\right) - \log_2 e \sum_{k=1}^{K_1} \boldsymbol{\gamma}_{i_k}^T \mathbf{G}_{i_k} \boldsymbol{\psi}_{i_k}
 \end{array}
\end{sequation}%
where
\begin{sequation}\label{eq:mutual_info}
\begin{array}{l}
I\left( {{\bf{d}}_{i_k } ;{\bf{z}}_{i_k } \left| {\sqrt {{\bf{T}}_{i_k } } {\bf{B}}_{i_k } } \right.} \right) = \log _2 M_{i_k } - \\ \frac{1}{{M_{i_k } }}\! \sum\limits_{m = 1}^{M_{i_k } } {E_{{\bf{v}} } \!  \left\{ \!  {\log _2 \sum\limits_{p = 1}^{M_{i_k } } {e^{ - \left( {\left\| {\sqrt {{\bf{T}}_{i_k } } {\bf{B}}_{i_k } \left( {{\bf{a}}_{k,p}  - {\bf{a}}_{k,m} } \right) + {\bf{v}} } \right\|^2  - \left\| {{\bf{v}} } \right\|^2 } \right)} } } \! \right\} \! }
\end{array}
\end{sequation}%

Eq. (\ref{eq:GAUMutuall_2}) implies that, in the large-system regime,
the mutual information $I\left( {{\bf{d}}_{\cal A} ;{\bf{y}}\left| {{\bf{d}}_{{{\cal A}^c} } } \right.} \right)$
can be evaluated by calculating the sum of the mutual informations of all individual users over the equivalent
channel ${\bf{T}}_{i_k }, k = 1,2,\cdots,K_1$.  Compared to the conventional method of calculating mutual information
which requires a search over all possible combinations of all users' signal sets \cite{Wang2011},
the asymptotic expression in Proposition \ref{prop:ach_rate_mac} has a significantly lower implementation
complexity. Moreover, given
statistical channel knowledge (i.e., $\{\mathbf{U}_{{\rm T}_k}\}_{\forall k}, \{\mathbf{U}_{{\rm R}_k}\}_{\forall k}, \{\mathbf{G}_k\}_{\forall k}$), the asymptotic mutual information
can be obtained from Proposition \ref{prop:ach_rate_mac}, without knowing the actual channel realization. Thus, the derived asymptotic expression can be used to design
transceivers which only require knowledge of the channel statistics,  see Section \ref{sec:Precoding}.

\subsection{WSR Problem}
It is well known that the capacity region
of the MIMO MAC $(R_1, R_2, \cdots, R_K)$ can be achieved by solving the WSR optimization problem \cite{goldsmith2003capacity}.
Without loss of generality, assume  weights $\mu _1 \geq \mu _2 \geq \cdots \geq  \mu _K \geq \mu _{K+1} = 0$, i.e.,
 users are decoded in the order $K, K -1, \cdots, 1$ \cite{Wang2011}.  Then, the WSR problem can be expressed as
\begin{sequation}
\begin{array}{l}
    R_{\rm sum} ^ {w} \left({\mathbf{B}}_1 ,{\mathbf{B}}_2 , \cdots, {\mathbf{B}}_K \right) =  \\
    \hspace{2cm}  \mathop {\max }\limits_{{\mathbf{B}_1, \mathbf{B}_2, \cdots, \mathbf{B}_K}}
\sum_{k = 1}^{K} \Delta_{k} f (\mathbf{B}_{1}, \mathbf{B}_{2}, \cdots, \mathbf{B}_{k}) \label{eq:obj_wsr_2}
\end{array}
\end{sequation}%
\begin{sequation}
       \text{tr} \left(\mathbf{B}_{k}\mathbf{B}^H_{k}\right) \leq P_{k}, \quad k = 1, 2, \cdots, K \label{eq:st_power}
\end{sequation}%
where $\Delta_{k} = \mu _k - \mu _{k+1}$, $k = 1,2,\cdots,K$. $f (\mathbf{B}_{1}, \mathbf{B}_{2}, \cdots, \mathbf{B}_{k}) = I(\mathbf{d}_1,
\cdots, \mathbf{d}_k ; \mathbf{y}|\mathbf{d}_{k+1},\cdots,\mathbf{d}_{K})$ can be evaluated based on
Proposition \ref{prop:ach_rate_mac}.
When $\mu_{1} = \mu_{2} = \cdots = \mu_{K} = 1$, (\ref{eq:obj_wsr_2}) reduces to the sum-rate maximization.

\section{Linear Precoding Design for MIMO MAC}\label{sec:Precoding}
\subsection{Necessary Conditions for Asymptotically Optimal Precoders}
\begin{prop}\label{prob:nec_cod_mac}
The asymptotically optimal precoders for  maximization of
the WSR in (\ref{eq:obj_wsr_2}) satisfy the following conditions
\begin{sequation}\label{eq:nec_cod_1_asy}
\begin{array}{l}
\kappa _l {\bf{B}}_l \! = \! \log _2 e \ \sum\limits_{k = l}^K {\Delta _k \left( {\sum\limits_{t = l}^k {\left( {{\boldsymbol{\Theta }}_{1,k,t,l} \! - \! {\boldsymbol{\Theta }}_{2,k,t,l} } \right)}  \!+ \! {\boldsymbol{\Theta }}_{3,k,l} } \right)}, \\
\hspace{4cm} l = 1,2,\cdots,K
\end{array}
\end{sequation}%
\begin{sequation}\label{eqn:nec_cod_2_asy}
\kappa_l \left( { {{\rm{tr}}\left( {{\bf{B}}_l^H {\bf{B}}_l } \right) - P_l} } \right) = 0, \  l = 1,2,\cdots,K
\end{sequation}%
\begin{sequation}\label{eqn:nec_cod_3_asy}
{ {{\rm{tr}}\left( {{\bf{B}}_l^H {\bf{B}}_l } \right) - P_l} } \leq 0, \  l = 1,2,\cdots,K
\end{sequation}%
\begin{sequation}\label{eqn:nec_cod_4_asy}
\kappa_l \geq 0,   \  l = 1,2,\cdots,K
\end{sequation}%
where ${\boldsymbol{\Theta}}_{1,k,t,l} \in \mathbb{C}^{N_t \times N_t} $ are matrices with elements
\begin{sequation}\label{eqn:Theta_mn}
\begin{array}{l}
\left[ {{\boldsymbol{\Theta }}_{1,k,t,l} } \right]_{mn}  = {\rm{tr}}\left( {{\boldsymbol{\Omega }}^{(k)}_{t} {\bf{B}}_t^H \sqrt { \left({\bf{T}}^{(k)}_{t}\right)^H } {\bf{D}}_{k,t,l,mn} } \right), \\
\hspace{4cm}  m,n = 1,2,\cdots,N_t
\end{array}
\end{sequation}%
\begin{sequation}\label{eqn:Theta_2}
\left[ {{\boldsymbol{\Theta }}_{2,k,t,l} } \right]_{mn}  =  - {\boldsymbol{\omega }}_{k,t,l,mn}^T {\bf{G}}_t {\boldsymbol{\psi }}_{k,t}  + \left({\boldsymbol{\gamma }^{(k)}_{t}}\right)^T {\bf{G}}_t {\boldsymbol{\theta }}_{k,l,mn} \sigma \left( {k - l} \right)
\end{sequation}%
\begin{sequation}\label{eqn:Theta_3}
\left[ {{\boldsymbol{\Theta }}_{3,k,l} } \right]_{mn}  = {\rm{tr}}\left( {\left( {{\bf{I}}_{N_r} + {\bf{R}}_{A_k } } \right)^{ - 1} {\bf{L}}_{k,l,mn} } \right).
\end{sequation}
\vspace{-0.1cm}
with
\begin{sequation}\label{eqn:D_mn}
\begin{array}{l}
{\bf{D}}_{k,t,l,mn}  =  - \frac{1}{2}{\bf{U}}_{{\rm{T}}_t } {\rm{diag}} \left( {{\bf{G}}_t^T {\boldsymbol{\gamma }}^{(k)}_{t} } \right)^{ - 1/2}  \\
 \hspace{-0.5cm} \times {\rm{diag}} \left( {{\bf{G}}_t^T {\boldsymbol{\omega }}_{k,t,l,mn} } \right) {\bf{U}}_{{\rm{T}}_t }^H {\bf{B}}_t  + \sqrt {\left({\bf{T}}^{(k)}_{t}\right)^H } {\bf{e}}_m {\bf{e}}_n^H \sigma \left( {t - l} \right)
\end{array}
\end{sequation}%
\begin{sequation}\label{eqn:omega_mn}
\begin{array}{l}
 {\boldsymbol{\omega }}_{k,t,l,mn}  \! = \!
 \! \left[ \! {\bf{u}}_{{\rm{R}}_t ,1}^H \!\left( {{ \mathbf{I}_{N_r}\!+\! \mathbf{R}}_{{\cal {A}}_k } } \right)^{ - 1}\! {\bf{L}}_{k,l,mn} \!\left( {  { \mathbf{I}_{N_r}\! + \!\mathbf{ R}}_{{\cal {A}}_k } } \right)^{ - 1}\! {\bf{u}}_{{\rm{R}}_t ,1} ,   \right. \\  {\bf{u}}_{{\rm{R}}_t ,2}^H \left( {{\mathbf{I}_{N_r}+ \mathbf{R}}_{{\cal {A}}_k } } \right)^{ - 1} {\bf{L}}_{k,l,mn} \left( {{\mathbf{I}_{N_r}+ \mathbf{R}}_{{\cal {A}}_k } } \right)^{ - 1} {\bf{u}}_{{\rm{R}}_t ,2} , \cdots , \\
 \left. {\bf{u}}_{{\rm{R}}_t ,N_r }^H \left( {{\mathbf{I}_{N_r}+ \mathbf{R}}_{ {\cal {A}}_k } } \right)^{ - 1} {\bf{L}}_{k,l,mn} \left( {{\mathbf{I}_{N_r}+ \mathbf{R}}_{{\cal {A}}_k } } \right)^{ - 1} {\bf{u}}_{{\rm{R}}_t ,N_r } \right]^T  \\
 \end{array}
\end{sequation}%
\begin{sequation}\label{eqn:L_mn}
{\bf{L}}_{k,l,mn}  = {\bf{U}}_{{\rm{R}}_l } {\rm{diag}} \left( {{\bf{G}}_l {\boldsymbol{\theta }}_{k,l,mn} } \right){\bf{U}}_{{\rm{R}}_l }^H
\end{sequation}%
\begin{sequation}\label{eqn:theta_mn}
\begin{array}{l}
{\boldsymbol{\theta }}_{k,l,mn} =  \left[ {{\bf{u}}_{{\rm{T}}_l ,1}^H {\bf{Q}}_{k,l,mn} {\bf{u}}_{{\rm{T}}_l ,1} ,{\bf{u}}_{{\rm{T}}_l ,2}^H {\bf{Q}}_{k,l,mn} {\bf{u}}_{{\rm{T}}_l ,2}, \cdots, } \right. \\
\hspace{4cm} \left.{{\bf{u}}_{{\rm{T}}_l ,N_t }^H {\bf{Q}}_{k,l,mn} {\bf{u}}_{{\rm{T}}_l ,N_t } } \right]^T
 \end{array}
\end{sequation}%
\begin{sequation}\label{Q_nec}
{\bf{Q}}_{k,l,mn}  = {\bf{B}}_l {\boldsymbol{\Delta }}_{k,l,mn} {\bf{B}}_l^H  + {\bf{B}}_l {\bf{E}}_{k,l} {\bf{e}}_n {\bf{e}}_m^H.
\end{sequation}%
Here, ${\bf{e}}_m$ is a unit-vector with the $m$th element being one
and all other elements zeros, and 
$\sigma \left[ x \right]$ denotes the Kronecker delta function where
$\sigma \left[ x \right] = 1$, $x =0$, and $\sigma \left[ x \right] = 0$, otherwise.
Also, $\mathbf{T}^{(k)}_{t}$, $\mathbf{R}^{(k)}_{t}$, ${\boldsymbol{\gamma}}^{(k)}_{t}$, ${\boldsymbol{\psi}}^{(k)}_{t}$,
and ${\boldsymbol{\Omega }}^{(k)}_{t}$ are obtained based on Definition \ref{Definition_1}
and (\ref{eq:mse}) by setting ${\cal{A}} = \left\{1,2,\cdots, k\right\}$, $t = 1,2,\cdots,k$.
Furthermore,  matrix $\mathbf{R}_{{\cal {A}}_k} \in \mathbb{C}^{N_r \times N_r} $ in (\ref{eqn:omega_mn})
is given by $\mathbf{R}_{{\cal {A}}_k} = \sum_{t=1}^{k} \mathbf{R}^{(k)}_{t}$.
Moreover, ${\boldsymbol{\Delta}}_{k,l,mn} \in \mathbb{C}^{N_t \times N_t}$ in (\ref{Q_nec}) is a matrix where
 element $\left[{\boldsymbol{\Delta}}_{k,l,mn} \right]_{ij}$ is taken from row
$p = i + (j-1) N_t$ and column $q= m + (n -1)N_t$ of matrix
${\boldsymbol{\Xi }}_{k,l} \in \mathbb{C}^{N_t^2 \times N_t^2}$, $1\leq i \leq N_t$, $1\leq j \leq N_t$, defined as
\begin{sequation}\label{eqn:Xi_mn}
\begin{array}{l}
 {\boldsymbol{\Xi }}_{k,l} = \\
 \hspace{-0.2cm}- E_{\mathbf{v}}\! \!\left[ \! \! E_{\mathbf{d}_{l}} \! \! \left[ \!  {{\bf{K}}_{N_t^2 } \! \left( \! {{\boldsymbol{\Phi }}_{k,{\bf{d}}_l {\bf{d}}_l^H } \! \otimes \!  \left[\! {{\boldsymbol{\Phi }}_{k,{\bf{d}}_l {\bf{d}}_l^H }^T {\bf{B}}_l^T \! \sqrt {\left({\bf{T}}^{(k)}_{l}\right)^T }  \sqrt {\left({\bf{T}}^{(k)}_{l}\right)^* } } \! \right]} \!\right)}\! \!\right] \!\! \right] \\
  \hspace{-0.2cm} -E_{\mathbf{v}} \! \left[  \!  E_{\mathbf{d}_{l}}  \left[ \! { \!\left( \! {{\boldsymbol{\Psi }}_{k,{\bf{d}}_l {\bf{d}}_l^T }^*  \otimes \left[ {{\boldsymbol{\Psi }}_{k,{\bf{d}}_l {\bf{d}}_l^T }
{\bf{B}}_l^T \sqrt {\left({\bf{T}}^{(k)}_{l}\right)^T } \sqrt {\left({\bf{T}}^{(k)}_{l}\right)^* } } \right]} \!\right)}\! \right] \!\right]. \\
 \end{array}
\end{sequation}%
Here, ${\bf{K}}_{N_t^2 } \in \mathbb{C}^{N_t^2 \times N_t^2} $ denotes a communication matrix \cite{payaro2009hessian},
and
\begin{sequation}\label{eqn:Phi}
{\boldsymbol{\Phi }}_{k,{\bf{d}}_l {\bf{d}}_l^H }  = \left( {{\bf{d}}_l  -{\hat{\bf{d}}^{(k)}_{l} }} \right)\left( {{\bf{d}}_l  -{\hat{\bf{d}}^{(k)}_{l} }} \right)^H
\end{sequation}%
\begin{sequation}\label{eqn:Psi}
{\boldsymbol{\Psi }}_{k,{\bf{d}}_l {\bf{d}}_l^T }  = \left( {{\bf{d}}_l  -{\hat{\bf{d}}^{(k)}_{l} }} \right)\left( {{\bf{d}}_{l}  -{\hat{\bf{d}}^{(k)}_{l} }} \right)^T
\end{sequation}%
\begin{sequation}\label{eqn:Ekl}
{\bf{E}}_{k,l} {\rm{ = }}E_{\mathbf{v}} \left[  E_{\mathbf{d}_{l}}\left[{\left( {{\bf{d}}_l  - \hat{\bf{d}}^{(k)}_{l} } \right)\left( {{\bf{d}}_l  - \hat{\bf{d}}^{(k)}_{l} } \right)^H \left| {{\bf{z}}^{(k)}_{l} } \right.} \right] \right].
\end{sequation}%
Vectors $\mathbf{z}^{(k)}_{l}$ and
$\mathbf{\hat{d}}^{(k)}_{l}$ of the $l$th user are obtained based on (\ref{eq:EqScalGAUEach}), (\ref{eq:hatx_k}),
and $\mathbf{T}^{(k)}_{l}$, $l = 1,2,\cdots,k$. 
\begin{proof}
In order to solve the WSR optimization problem in (\ref{eq:obj_wsr_2}), we
can establish a Lagrangian cost function for the precoding matrices.
Then, based on the KKT conditions and the matrix derivation technique \cite{Hjorungnes2011}, we can obtain Proposition \ref{prob:nec_cod_mac}.
Due to the space limitation, details of the proof are omitted here,
 and will be given in an extended journal version of this paper.
\end{proof}
\end{prop}

\subsection{Iterative Algorithm for Weighted Sum Rate Maximization}
The necessary condition in (\ref{eq:nec_cod_1_asy}) indicates that the precoding matrices of different
users depend on each another.  Thus, 
the optimal precoding matrices ${\bf{B}}_l$,  $l = 1,2,\cdots, K$ have to be 
found numerically. Problem (\ref{eq:obj_wsr_2}) is a
 multi-variable optimization problem. Therefore,
we employ the alternating optimization method which iteratively updates one precoder
at a time with the other precoders being fixed. This is a commonly used  approach in handling
multi-variables optimization problems \cite{Wang2011}. In each iteration step, we optimize the precoders
along the gradient descent direction which corresponds to the partial derivative of the WSR
(\ref{eq:obj_wsr_2}) with respect to $\mathbf{B}_l$, $l = 1,2,\cdots, K$. The
partial derivative is given  by the right

\begin{alg} \label{Gradient_MAC}
Gradient descent algorithm for  WSR
maximization with respect to $\left\{{\mathbf{B}}_1,{\mathbf{B}}_2 ,\cdots, {\mathbf{B}}_K \right\}$
\vspace*{1.5mm} \hrule \vspace*{1mm}
  \begin{enumerate}

\itemsep=0pt

\item Initialize ${\mathbf{B}}_{l}^{(1)}$, $l = 1,2,\cdots,K$, with ${\tr\left( {\left({{\bf{B}}_{l}^{(1)}} \right)^H {\bf{B}}_{l}^{(1)}} \right)} = P_l$, $l= 1,2,\cdots,K$. Set initialization index to $n = 1$. Initialize ${\boldsymbol{\gamma }}_{t}^{(k), \left( {{0}} \right)}$ and ${\boldsymbol{\psi }}_{t}^{(k), \left( {{0}} \right)}$, $t = 1,2,\cdots,k$, $k = 1,2,\cdots,K$. Set the tolerance $\varepsilon$ and the maximum iteration number
    $N_{\rm max}$. Select values for the backtracking line search  parameters $\theta$ and $\omega$ with $\theta \in (0,0.5)$ and $\omega  \in (0,1)$.

\item Using Definition \ref{Definition_1}, compute ${\bf{T}}^{(k)}_{t}$, ${\bf{R}}^{(k)}_{t}$, ${\boldsymbol{\gamma }}_{t}^{(k),\left( {{n}} \right)}$, and ${\boldsymbol{\psi }}_{t}^{(k),\left( {{n}} \right)}$  for  ${\bf{B}}_k^{\left( n \right)}$, ${\boldsymbol{\gamma }}_{t}^{(k), \left( {{n-1}} \right)}$, and ${\boldsymbol{\psi }}_{t}^{(k),\left( {{n-1}} \right)}$, $t = 1,2,\cdots,k$, $k = 1,2,\cdots,K$.

\item Using (\ref{eq:obj_wsr_2}), compute the  asymptotic value $R_{\rm sum,asy} ^ {w,(n)} \left({\mathbf{B}}_1 ,{\mathbf{B}}_2 , \cdots, {\mathbf{B}}_K \right)$  for  ${\bf{B}}_k^{\left( n \right)}$,
${\bf{T}}^{(k)}_{t}$, ${\bf{R}}^{(k)}_{t}$, ${\boldsymbol{\gamma }}_{t}^{(k),\left( {{n}} \right)}$, and ${\boldsymbol{\psi }}_{t}^{(k), \left( {{n}} \right)}$, $t = 1,2,\cdots,k$, $k = 1,2,\cdots,K$.

\item Using (\ref{eq:nec_cod_1_asy}),  compute  the asymptotic gradient
$\nabla_{\mathbf{B}_l} R_{\rm sum, asy} ^ {w, (n)} \left({\mathbf{B}}_1 ,{\mathbf{B}}_2 , \cdots, {\mathbf{B}}_K \right) $, $l = 1,2,\cdots,K$, for ${\bf{B}}_k^{\left( n \right)}$,
${\bf{T}}^{(k)}_{t}$, ${\bf{R}}^{(k)}_{t}$, ${\boldsymbol{\gamma }}_{t}^{(k), \left( {{n}} \right)}$, and ${\boldsymbol{\psi }}_{t}^{(k), \left( {{n}} \right)}$, $t = 1,2,\cdots,k$, $k = l,l + 1,\cdots,K$.

\item Set $l: = 1$.

\item Set the step size $u:= 1$.

\item Evaluate $c = \alpha u \left\| {\nabla _{{\bf{B}}_l } }  R_{\rm sum, asy} ^ {w, (n)}
\left({\mathbf{B}}_1 ,{\mathbf{B}}_2 , \cdots, {\mathbf{B}}_K \right) \right\|_F ^2$.
If $c$ is smaller than a threshold, then go to step $13$.

\item Compute $\widetilde{\mathbf{B}}_l^{\left( n \right)}  = {\mathbf{B}}_l^
{\left( n \right)} + u {\nabla _{{\bf{B}}_l } }  R_{\rm sum, asy} ^ {w,(n)}\left({\mathbf{B}}_1 ,{\mathbf{B}}_2 , \cdots, {\mathbf{B}}_K \right)$.

\item If ${\tr\left(  \left({ \widetilde{\bf{B}}_l^{\left( n \right)} }\right)^H { \widetilde{\bf{B}}_l^{\left( n \right)} } \right)} > P_l$,
update ${\bf{B}}_l^{\left( {n + 1} \right)}  = \frac{{\sqrt {P_l } \widetilde{\bf{B}}_l^{\left( n \right)} }}{{ { {\left\| {\widetilde{\bf{B}}_l^{\left( n \right)} } \right\|_F } } }}$; otherwise, ${\bf{B}}_l^{\left( {n + 1} \right)} = \widetilde{\bf{B}}_l^{\left( n \right)}$.

\item Using Definition \ref{Definition_1}, compute ${\bf{T}}^{(k)}_{t}$, ${\bf{R}}^{(k)}_{t}$, ${\boldsymbol{\gamma }}_{t}^{(k), \left( {{n}} \right)}$, and ${\boldsymbol{\psi }}_{t}^{ (k), \left( {{n}} \right)}$ for ${\bf{B}}_1^{\left( {n + 1} \right)} , \cdots ,{\bf{B}}_l^{\left( {n + 1} \right)}$,
${\bf{B}}_{l + 1}^{\left( n \right)} , \cdots ,{\bf{B}}_K^{\left( n \right)}$, ${\boldsymbol{\gamma }}_{t}^{(k), \left( {{n-1}} \right)}$,
and ${\boldsymbol{\psi }}_{t}^{(k),\left( {{n -1}} \right)}$, $t = 1,2,\cdots,k$, $k = 1,2,\cdots,K$.

\item Using (\ref{eq:obj_wsr_2}), compute $R_{\rm sum, asy} ^ {w,(n + 1)}\left({\mathbf{B}}_1 ,{\mathbf{B}}_2 , \cdots, {\mathbf{B}}_K \right)$ for ${\bf{B}}_1^{\left( {n + 1} \right)} , \cdots ,{\bf{B}}_l^{\left( {n + 1} \right)}$,
${\bf{B}}_{l + 1}^{\left( n \right)} , \cdots ,{\bf{B}}_K^{\left( n \right)}$, ${\bf{T}}^{(k)}_{t}$, ${\bf{R}}^{(k)}_{t}$, ${\boldsymbol{\gamma }}_{t}^{(k),\left( {{n}} \right)}$, and ${\boldsymbol{\psi }}_{t}^{(k),\left( {{n}} \right)}$, $t = 1,2,\cdots,k$, $k = 1,2,\cdots,K$.

\item Set $u: = \beta u$.  If $R_{\rm sum, asy} ^ {w, (n + 1)} < R_{\rm sum, asy} ^ {w, (n )} + c $, go to step $7$.

\item If $l \leq K$,  $l: = l + 1$,  go to step $6$.

\item If $R_{\rm sum, asy} ^ {w, (n + 1)}  - R_{\rm sum, asy} ^ {w, (n )} > \varepsilon$ and $n < N_{\rm max}$, set $n: = n + 1$, go to step 2; otherwise, stop the algorithm.

 \vspace*{1mm} \hrule

  \end{enumerate}
\end{alg}
\null
\par

\noindent  hand side of (\ref{eq:nec_cod_1_asy}).
The backtracking line search method is incorporated to determine the step size for each gradient update \cite{Boyd2004}. 
In addition, if the updated precoder exceeds the power constraint $\tr\left\{{{\mathbf{B}}}_l{{\mathbf{B}}}_l^H\right\} > P_l$, we  project $\mathbf{{B}}_l$ onto the feasible set through a normalization step: ${{\mathbf{B}}}_l: = \sqrt{P_l} {\mathbf{{B}}}_l /\sqrt{{{\tr\left\{{\mathbf{{B}}_l {\mathbf{B}}}_l^H\right\}}}}$ \cite{Palomar2006TIT}.
The resulting algorithm is given in Algorithm \ref{Gradient_MAC}.

The computational complexity of  linear precoder design algorithms  for the MIMO MAC with finite alphabet inputs
is determined by the required number of  summations in calculating the mutual information and the MSE matrix
(e.g., (\ref{eq:mutual_info}), (\ref{eqn:Xi_mn}) or \cite[Eq. (5)]{Wang2011}, \cite[Eq. (24)]{Wang2011}). The conventional precoder
design for instantaneous CSI in \cite{Wang2011} requires summations over all possible transmit
vectors of all users. For this reason, the computational complexity of the conventional precoding
design scales linearly with $\left( {\prod\nolimits_{k = 1}^K {Q_k } } \right)^{2 N_t }$. However, (\ref{eq:GAUMutuall_2})
and (\ref{eq:nec_cod_1_asy}) imply that Algorithm \ref{Gradient_MAC} only requires summations over each user's own
possible transmit vectors to design the precoders.  Accordingly, the  computational complexity of the proposed Algorithm \ref{Gradient_MAC} for statistical CSI
grows linearly with $\sum\nolimits_{k = 1}^K {Q_k^{2 N_t } }$.
As a result,
the computational complexity of Algorithm \ref{Gradient_MAC} is  several orders of magnitude lower than that of the conventional design. To exemplify this
more clearly,  we give an example. We consider a practical massive MIMO MAC system where the base station is equipped with a large number of antennas and
serves multiple users having much smaller numbers of antennas. In particular, we assume $N_r = 64$, $N_t = 4$, $K = 4$,
$\mu_1 = \mu_2 = \mu_3 = \mu_4$, and all users
employ the same modulation constellation.  The numbers  of summations required for calculating the mutual information and the MSE matrix
in Algorithm \ref{Gradient_MAC} and in the precoding
design in \cite{Wang2011} are listed in Table \ref{tab:mac_dim} for different modulation formats.

\begin{table}[!t]
\centering
\caption{Number of summations required for calculating the mutual information and the MSE matrix.} \label{tab:mac_dim}
\vspace*{1.5mm}
\begin{tabular}{|c|c|c|c|c|c|c|}
\hline
\  Modulation     &  QPSK & 8PSK &   16 QAM     \\ \hline
\  Algorithm \ref{Gradient_MAC}        &   262144  & 6.7 e+007  &  1.7 e+010    \\ \hline
 \  Design Method in \cite{Wang2011}  & 1.85  e+019  & 7.9 e+028  &  3.4 e+038       \\ \hline
\end{tabular}
\end{table}

We observe from Table \ref{tab:mac_dim} that Algorithm \ref{Gradient_MAC} significantly
reduces number of summations required for  MIMO MAC precoder design for finite alphabet inputs.
Moreover, since Algorithm \ref{Gradient_MAC} is based on the channel statistics
$\{\mathbf{U}_{{\rm T}_k}\}_{\forall k}$, $\{\mathbf{U}_{{\rm R}_k}\}_{\forall k}$, $\{\mathbf{G}_k\}_{\forall k}$,
it avoids the time-consuming averaging process over each channel realization for
the mutual information in (\ref{mutual_info_1}). In addition, Algorithm \ref{Gradient_MAC} is executed only once since
the precoders are constant as long as
the channel statistics do not change, whereas the algorithm in \cite{Wang2011} has to be executed for each channel realization.
Due to the non-convexity of the objective function
$R_{\rm sum} ^ {w} \left({\mathbf{B}}_1 ,{\mathbf{B}}_2 , \cdots, {\mathbf{B}}_K \right)$ in general,
Algorithm \ref{Gradient_MAC}  will find a local maximizer of the WSR. Therefore,
we run Algorithm 1 for several random initializations ${\mathbf{B}}_{k}^{(1)}$
and select the result that offers the maximal WSR as the final
design solution \cite{Wang2011,Wu2012TWC}.

\section{Numerical Results}
In this section, we provide examples to illustrate the performance
of the proposed iterative optimization algorithm. We consider a two-user
MIMO MAC system with two transmit antennas and two
receive antennas\footnote{Although the derivations in this paper are based on the assumption that $N_t$ and $N_r$ both approach infinity,
we want to show that the proposed Algorithm 1 can perform well even for a MIMO MAC system with a small number of antennas. Therefore, we consider an example of $N_t = N_r = 2$  in the simulations.} for each user. We assume equal user powers
$P_1 = P_2 = P$, $\mu_1 = \mu_2 = 1$, and the same modulation format for both users.
The average signal-to-noise ratio (SNR) for the MIMO MAC with statistical CSI at the transmitter is given by
${\rm{SNR}} = \frac{{E\left[ {{{\rm{tr}}\left( {{\bf{H}}_k {\bf{H}}_k^H } \right)}} \right]}P}{{N_t N_r }}$.

For illustrative purpose, we consider an example of
jointly correlated fading channel matrices for two users. The channel statistics in (\ref{H_channel}) are given by
\begin{sequation}\label{H_user_1}
\begin{array}{lll}
{\bf{U}}_{{\rm{T}}_1 }   & = & \left[ \begin{array}{l}
  - 0.7830  \qquad  \qquad \qquad 0.6196 + 0.0547j \\
  - 0.6196 + 0.0547j  \ \, - 0.7830  \\
 \end{array} \right] \nonumber \\
{\bf{U}}_{{\rm{R}}_1 }   & = & \left[ \begin{array}{l}
0.9513    \qquad  \qquad \ \, - 0.0364 + 0.3061j \\
0.0364 + 0.3061j  \quad \, 0.9513 \\
 \end{array} \right] \nonumber \\
{{ \bf\widetilde{G}}_1}   & = &  \left[ \begin{array}{l}
 1.8366  \quad   0.3979 \\
0.6122   \quad  0.3061 \\
 \end{array} \right]
  \end{array}
\end{sequation}%
and
\begin{sequation}\label{H_user_1}
\begin{array}{lll}
{\bf{U}}_{{\rm{T}}_2 }  &= &\left[ \begin{array}{l}
  - 0.9628   \qquad  \qquad \qquad  0.2683 - 0.0313j \\
  - 0.2683 - 0.0313j  \ \,  - 0.9628  \\
 \end{array} \right] \nonumber \\
{\bf{U}}_{{\rm{R}}_2 }  &= &\left[ \begin{array}{l}
0.7757    \qquad  \qquad \ \,    - 0.0479 - 0.6293j \\
 0.0479 - 0.6293j  \quad \, 0.7757  \\
 \end{array} \right] \nonumber \\
{{\bf\widetilde{G}}_2} & =& \left[ \begin{array}{l}
 0.1242  \quad  1.2415 \\
  0.1862  \quad  1.5519 \\
 \end{array} \right].
 \end{array}
\end{sequation}%

\begin{figure}[!t]
\centering
\includegraphics[width=0.5\textwidth]{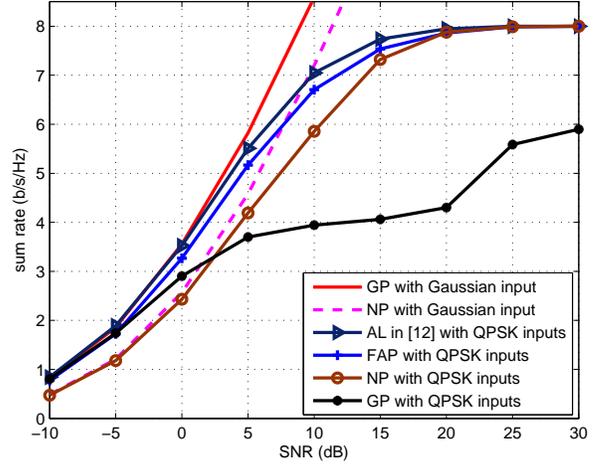}
\caption {\space\space Average sum-rate of two-user MIMO MAC with QPSK modulation.}
\label{Sum_Rate_QPSK_mac}
\end{figure}

Figure \ref{Sum_Rate_QPSK_mac} plots the sum-rate curves for different transmission schemes and
QPSK inputs. We employ the Gauss-Seidel algorithm together with stochastic programming
to obtain the optimal covariance matrices of both users under the Gaussian input assumption \cite{wen2011sum}.
Then, we decompose the obtained
optimal covariance matrices $\left\{ {{\mathbf{Q}}_1 , {\mathbf{Q}}_2, \cdots ,{\mathbf{Q}}_K } \right\}$
as ${\mathbf{Q}}_k = {\mathbf{U}}_k \boldsymbol{\Lambda} _k {\mathbf{U}}_k^H$, and set ${\mathbf{B}}_k
= {\mathbf{U}}_k \boldsymbol{\Lambda} _k^{\frac{1}{2}}$, $k = 1,2, \cdots, K$. Finally, we calculate
the average sum-rate of this precoding design under finite alphabet constraints. We denote the corresponding sum-rate
as ``GP with QPSK inputs". For the case without precoding, we set ${\mathbf{B}}_1
 = {\mathbf{B}}_2  = \sqrt{\frac{P}{N_t}} \mathbf{I}_{N_t}$.
 We denote the corresponding sum-rate as ``NP with QPSK inputs".
Also, the sum-rates achieved with the Gauss-Seidel algorithm  and without precoding for Gaussian inputs
 are also plotted in Figure \ref{Sum_Rate_QPSK_mac}, and denoted as ``GP with Gaussian input" and
``NP with Gaussian input", respectively. For comparison purpose, we plot the average sum rate achieved by Algorithm 1 in \cite{Wang2011}
 with instantaneous CSI, which is denoted as ``AL in [12] with QPSK inputs". We denote
 the proposed design in Algorithm 1 as ``FAP with QPSK inputs".
We observe from Figure \ref{Sum_Rate_QPSK_mac}
that for QPSK inputs, the proposed algorithm achieves a better sum-rate performance than
the other precoding schemes with statistical CSI. In particular, at a sum-rate of $4$ b/s/Hz,
the SNR gains of the proposed algorithm over the ``NP with Gaussian input" design and
the ``GP with Gaussian input" design are $2.5$ dB and $11$ dB, respectively.
The sum rate achieved by the proposed algorithm with statistical CSI is
close to the sum rate achieved by Algorithm 1 in \cite{Wang2011} with instantaneous CSI.
At a target sum rate of $4$ b/s/Hz, the SNR gap between the proposed algorithm and
Algorithm 1 in \cite{Wang2011} is less than 1 dB.  The sum-rates achieved by the ``GP with Gaussian input" design almost
remain unchanged for SNRs between $10$ dB and $20$ dB.
Similar to the point-to-point MIMO case \cite{zeng2012linear}, this is because the Gauss-Seidel algorithm design implements
a  ``water filling" power allocation policy within this SNR region.
As a result, when the SNR is smaller than a threshold (e.g., 20 dB in this case), the precoders allocate most
energy to the strongest subchannels and allocates little to the weaker subchannels. Therefore,
one eigenvalue of ${\mathbf{Q}}_k$ may approach zero. For finite alphabet inputs, this power allocation policy may result
in allocating most of the power to  subchannels
that are close to saturation. It will lead to a waste of transmission power and impede the
further improvement of the sum-rate performance.

\begin{figure}[!t]
\centering
\includegraphics[width=0.5\textwidth]{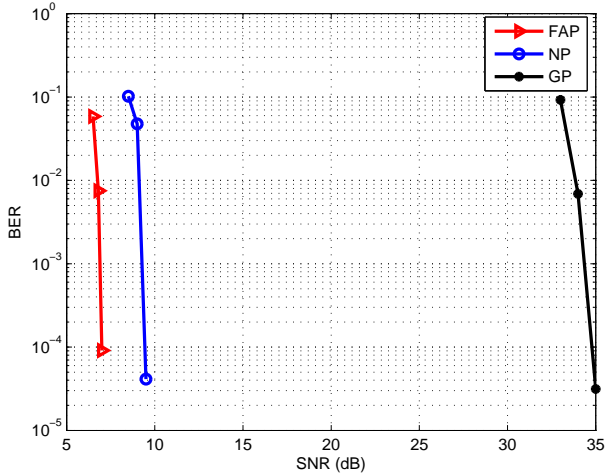}
\caption {\space\space BER of two-user MIMO MAC with QPSK modulation.}
\label{BER_QPSK_mac}
\end{figure}

Next, we verify
the performance of the proposed precoding design in a practical communication system. To this end, we employ
the low density parity check encoder and decoder simulation packages from \cite{Valenti}, with code rate $1/2$
and code length $L = 9600$. We employ the same transceiver structure as in \cite{Wang2011}.
Figure \ref{BER_QPSK_mac} depicts the average
coded BER performance of different precoding designs for QPSK inputs.
We observe that for a target
BER of $10^{-4}$, the proposed ``FAP" design achieves
$2.5$ dB SNR gain over the ``NP" design.
It is noted that code rate $1/2$ corresponds to a targeted
sum-rate of $4$ b/s/HZ for a two-user MIMO MAC system with two transmit antennas and QPSK inputs.
Therefore, the SNR gain for the coded BER  matches the SNR
gain for the sum-rate.  Also, for the coded BER, the ``FAP" design yields a $28$ dB SNR gain
over the ``GP" design, which
is larger than that for the sum-rate in Figure \ref{Sum_Rate_QPSK_mac}.
This is because for SNRs between $10$ dB and $20$ dB, the ``GP" design results in
a beamforming structure which allocates most power to the stronger subchannel.
Thus, the BER performance of the weaker subchannel is
much worse than that of the stronger subchannel.  Therefore,
the overall coded BER is high.

\section{Conclusion}
In this paper, we have studied the linear precoder design
for the $K$-user MIMO MAC with statistical CSI at the transmitter.
We formulated the problem from the standpoint of finite alphabet inputs based on a very general
jointly-correlated fading model.
We first obtained the WSR expression for a MIMO MAC system assuming a  jointly-correlated
fading model for the asymptotic large-system regime
under finite alphabet input constraints.
Then, we established
a set of necessary conditions for the precoding matrices which maximize the
asymptotic WSR. Subsequently, we proposed an iterative
algorithm to find the precoding matrices of all users with statistical CSI at the transmitter.  In the proposed algorithm, the search space for each user is
its own modulation set, which significantly reduces the dimension
of the search space compared to a previously proposed precoding design method for
MIMO MAC with finite alphabet inputs and instantaneous CSI at the transmitter.
Numerical results showed that, for finite alphabet inputs, precoders designed with
the proposed iterative algorithm achieve substantial performance gains
over the precoders designed based on the Gaussian input assumption and transmissions without precoding.

\appendices

\section{Proof of Proposition \ref{prop:ach_rate_mac}}\label{sec:proof_ach_rate_mac}
Due to space limitations, we only outline the main steps leading to Proposition \ref{prop:ach_rate_mac}.
Details of the proof will be given in an extended journal version of this paper.
First, we consider the case $K_1 = K$. Define ${\bf{H}} = \left[ {{\bf{H}}_{1} \ {\bf{H}}_{2}  \cdots \mathbf{H}_{{K}} } \right]$,
$\mathbf{B} = {\rm{blockdiag}} \left\{ \mathbf{B}_{1},\mathbf{B}_{2}, \cdots, \right. $ $ \left. \mathbf{B}_{K} \right\}$,
$ \mathbf{x} = \left[ \mathbf{x}_{1}^T \ \mathbf{x}_{2}^T  \cdots \mathbf{x}_{K}^T \right]^T$, and
$ \mathbf{d} = \left[ \mathbf{d}_{1}^T \ \mathbf{d}_{2}^T  \cdots \mathbf{d}_{K}^T \right]^T$.
From (\ref{eq:Finite_Mutual}), the mutual information of the MIMO MAC
can be expressed as $ I(\mathbf{d}; {\bf y})= F - N_r \log_2 e $, where
$ F = -{E}_{{\bf y},\mathbf{H}}\left[\log_2 Z({\bf y},\mathbf{H})\right]$ and
$Z({\bf y},\mathbf{H}) = {E}_{\mathbf{x}}\left[e^{- \left\|{\bf y}- \mathbf{Hx}\right\|^2}\right]$. The expectations over $\mathbf{y}$ and $\mathbf{H}$ are difficult to perform because the
logarithm appears inside the average. The replica method, nevertheless, circumvents the difficulties by rewriting $F$ as
\begin{sequation} \label{eq:ap_sf_F}
F = -\log_2 e \lim_{r\rightarrow 0}\frac{\partial}{\partial r}\ln{ E}_{{\bf y},\mathbf{H}}\left[\left(Z({\bf y},\mathbf{H})\right)^r\right]
\end{sequation}%
The reformulation is very useful because it allows us to first evaluate ${E}_{{\bf y},\mathbf{H}}\left[\left(Z({\bf y},\mathbf{H})\right)^r \right]$ for an integer-valued $r$, before considering
 $r$ in the vicinity of $0$.
This technique is called the replica method \cite{Edwards1975}, and has been widely
adopted in the field of statistical physics \cite{Nishimori2001}.

Basically, to compute the expectation over $Z({\bf y},\mathbf{H})$, it is useful to introduce $r$ replicated signal vectors $\mathbf{x}_k^{(\alpha)}$, for $\alpha = 0, 1, \ldots,
r$, yielding
\begin{sequation} \label{eq:ap_sf_E1}
{ E}_{{\bf y},\mathbf{H}}\left[\left( Z({\bf y},\mathbf{H})\right)^r\right]={ E}_{\mathbf{H},\mathbf{X}}\left[\int \prod_{\alpha=0}^re^{- \left\|{\bf y}-\sum_{k=1}^{K} \mathbf{H}_k \mathbf{x}_k^{(\alpha)}\right\|^2} d \mathbf{y} \right]
\end{sequation}%
where $\mathbf{X} = \left[\mathbf{X}_1^T \, \mathbf{X}_2^T \, \cdots \, \mathbf{X}_K^T \right]^T$,
$\mathbf{X}_k = \left[ \mathbf{x}_k^{(0)} \,\mathbf{x}_k^{(1)}\, \cdots \,
\mathbf{x}_k^{(r)} \right]$, and $\{\mathbf{x}_k^{(\alpha)}\}$ are i.i.d. with distribution $p(\mathbf{x}_k)$. Now, the expectation over $\mathbf{y}$ can be performed because it is reduced to the Gaussian integral.
However, the expectation over $\mathbf{H}$ involves interactions among the replicated signal vectors. Define a set of random
vectors: $\mathbf{V} = \left[ \mathbf{V}_1 \, \mathbf{V}_2 \, \cdots \, \mathbf{V}_K \right]$,
$\mathbf{V}_k = \left[ \mathbf{v}_{k,1}^T \, \mathbf{v}_{k,2}^T \,\cdots \mathbf{v}_{k,N_r}^T \right]^T$, $\mathbf{v}_{k,n} =
\sum_{m}\mathbf{v}_{k,n,m}$,
$\mathbf{v}_{k,n,m} = \left[v_{k,n,m}^{(0)} \, v_{k,n,m}^{(1)} \, \cdots \, v_{k,n,m}^{(r)} \right]$, and
$v_{k,n,m}^{(\alpha)} = [\mathbf{W}_k]_{n,m} [\tilde{\mathbf{G}}_k]_{n,m} \mathbf{u}_{{\rm T}_k,m}^{H}
\mathbf{x}_k^{(\alpha)}$ for $\alpha=0,\dots,r$.
For given $\mathbf{X}_k$, $\mathbf{v}_{k,n,m}$ is a Gaussian random vector with zero mean and
covariance $\mathbf{Q}_{k,n,m} $, where $\mathbf{Q}_{k,n,m} \in {\mathbb C}^{(r+1)\times(r+1)}$
is a matrix with entries  $[\mathbf{Q}_{k,n,m}]_{\alpha \beta}={ E}_{w_{k,n,m}} \left[\left(v_{k,n,m}^{(\alpha)}\right)^H v_{k,n,m}^{(\beta)} \right]=g_{k,n,m}
\left(\mathbf{x}_k^{(\alpha)}\right)^H \mathbf{u}_{{\rm T}_k,m} \mathbf{u}_{{\rm T}_k,m}^{H} \mathbf{x}_k^{(\beta)}$,
$\forall \alpha, \beta$. For ease of notation, we further define $ \mathbf{T}_{k,m} = \mathbf{u}_{{\rm T}_k,m} \mathbf{u}_{{\rm T}_k,m}^{H}$
and $\mathbf{R}_{k,n} = \mathbf{u}_{{\rm R}_k,n}\mathbf{u}_{{\rm R}_k,n}^{H}$.
Therefore, we have $[\mathbf{Q}_{k,n,m}]_{\alpha \beta} = g_{k,n,m} \left(\mathbf{x}_k^{(\alpha)}\right)^H \mathbf{T}_{k,m}\mathbf{x}_k^{(\beta)}$.
Let ${\mathbb Q} = \{\mathbf{Q}_{k,n,m}\}_{\forall k,n,m}$, where ${\forall k,n,m}$ stands for $k = 1,2,\cdots,K$,
$m = 1,2,\cdots,N_t$, and $n = 1,2,\cdots,N_r$. It is useful to separate the expectation over $\mathbf{X}$ in (\ref{eq:ap_sf_E1}) into the expectation over ${\mathbb Q}$, and
then all possible $\mathbf{x}_k^{(\alpha)}$ configurations for a given ${\mathbb Q}$ by introducing a $\delta$-function,
\begin{sequation}\label{eq:ap_sf_E2}
{ E}_{{\bf y},\mathbf{H}}\left[\left(Z({\bf y},\mathbf{H},\sigma)\right)^r\right]=\int e^{ \mathcal{S}({\mathbb Q})}d\mu({\mathbb Q})
\end{sequation}%
where
\begin{sequation}
\begin{array}{l}
 \mathcal{S}({\mathbb Q}) = \\
  \hspace{-0.2cm}\ln \int  { E}_{\mathbf{V}}\left[\!\prod_{\alpha=0}^{r}e^{-\left\|{\bf y}- \sum_{k=1}^{K} \left(\sum_{n=1}^{N_r} \left(\sum_{m=1}^{K N_t} v_{k,n,m}^{(\alpha)} \right) \mathbf{u}_{{\rm R}_k,n} \right) \right\|^2}\! \right] d \mathbf{y}
\end{array}
 \end{sequation}%
\[ \mu({\mathbb Q}) \!= \! { E}_{\mathbf{X}}\left[\prod_{k,n,m}\prod_{0\leq\alpha\leq\beta}^{r}\delta\left(g_{k,n,m}  \left(\mathbf{x}_k^{(\alpha)}\right)^H \mathbf{T}_{k,m} \mathbf{x}_k^{(\beta)}  \right. \right. \]
\begin{sequation}
\begin{array}{l}
\hspace{1.5cm} \left. \left.- [\mathbf{Q}_{k,n,m}]_{(\alpha,\beta)}\right)\right]
\end{array}
\end{sequation}%
Using the inverse Laplace transform of
the $\delta$-function,  we can show that if $N_t$ is large, then $\mu({\mathbb Q})$ is dominated by the exponent term as
\begin{sequation}\label{eq:defJ}
\begin{array}{l}
    \mathcal{J}({\mathbb Q}) = \max_{\tilde{{\mathbb Q}}}  \left\{ \sum_{k,n,m} \tr \left(\tilde{\mathbf{Q}}_{k,n,m}\mathbf{Q}_{k,n,m}\right) \right.\\
\hspace{1cm}   \left.  -\ln{ E}_{\mathbf{X}}\left[e^{\sum_{k,m}\tr \left( \sum_{n} g_{k,n,m} \tilde{\mathbf{Q}}_{k,n,m}\mathbf{X}_k^H \mathbf{T}_{k,m}\mathbf{X}_k. \right)  }\right] \right\}
\end{array}
\end{sequation}%
We define the set $\tilde{{\mathbb Q}} =\{\tilde{\mathbf{Q}}_{k,n,m}\}_{\forall k,n,m}$ and $\tilde{\mathbf{Q}}_{k,n,m}\in {\mathbb C}^{(r+1)\times(r+1)}$ is a Hermitian matrix.
As a result, by applying the method of steepest descent to (\ref{eq:ap_sf_E2}), we have \cite{Tanaka2002TIT,Guo2005TIT}
\begin{sequation}\label{eq:calF1}
{\cal F} = - \lim_{N_t\rightarrow\infty}\ln { E}_{{\bf y},\mathbf{H}}\left[ \left(Z ({\bf y},\mathbf{H})\right)^r \right] \approx - \max_{{\mathbb Q}}\left\{{\cal S}({\mathbb Q})- {\cal J}({\mathbb Q}) \right\}
\end{sequation}%
The extremum over $\tilde{{\mathbb Q}}$  and  ${\mathbb Q}$ in (\ref{eq:defJ}) and (\ref{eq:calF1}) can be obtained via the saddle point method, yielding a set of self-consistent
equations. To avoid searching for the saddle-points over all possible ${\mathbb Q}$ and $\tilde{{\mathbb Q}}$,
we make the following {\it replica symmetry} (RS)
assumption for the saddle point:
\begin{sequation}
\mathbf{Q}_{k,n,m} = q_{k,n,m}{\bf 11}^H +(c_{k,n,m}-q_{k,n,m})\mathbf{I}_{r+1}
\end{sequation}%
\begin{sequation}
\tilde{\mathbf{Q}}_{k,n,m} =\tilde{q}_{k,n,m}{\bf 11}^H+(\tilde{c}_{k,n,m}-\tilde{q}_{k,n,m})\mathbf{I}_{r+1}
\end{sequation}%
where ${\bf{1}} \in \mathbb{C}^{ (r + 1) \times 1}$ is a vector with all elements equalling to one.
This RS assumption has been widely accepted in physics \cite{Nishimori2001}, and was also used in
communications \cite{wen2011sum,wen2007asymptotic,Tanaka2002TIT,RMuller2008JSAC}.

After some tedious algebraic manipulations, we obtain the RS solution of ${\cal F}$ as
\begin{sequation} \label{eq:midFree}
    {\cal F}=-\lim_{r\rightarrow 0}\frac{\partial}{\partial r}\max_{ \left\{c_{k,n,m}\right\},\left\{ {q}_{k,n,m}\right\}} \min_{\left\{\tilde{c}_{k,n,m} \right\},\left\{\tilde{{q}}_{k,n,m}\right\}}{\cal T}^{(r)}
\end{sequation}%
where
\begin{eqnarray*} \label{eq:TreplicaSym}
\begin{small}
\begin{array}{l}
    -{\cal T}^{(r)}
    = \int  {E}_{\mathbf{X}}\left[e^{ -\left\| \mathbf{z} - \sqrt{\boldsymbol{\Xi}'}\mathbf{x} \right\|^2 + \mathbf{x}^{H}(\boldsymbol{\Xi}'-\boldsymbol{\Xi})\mathbf{x} }\right] \\
\hspace{2cm}  \times  \left( {E}_{\mathbf{X}}\left[e^{ \left( \sqrt{\boldsymbol{\Xi}'}\mathbf{x} \right)^{H}\mathbf{z}
 \quad   + \mathbf{z}^{H}\left( \sqrt{\boldsymbol{\Xi}'}\mathbf{x} \right) - \mathbf{x}^{H} \boldsymbol{\Xi} \mathbf{x} } \right] \right)^r d\mathbf{z} \\
 \end{array}
 \end{small}
 \end{eqnarray*}%
\begin{sequation} \label{eq:TreplicaSym_2}
\begin{array}{l}
   \quad   +r\ln\det\left(\mathbf{I}_{N_r}+\sum_{k,n} \left(\sum_{m}c_{k,n,m}-q_{k,n,m}\right) \mathbf{R}_{k,n} \right)  \\
   \quad  + N_r\ln(r+1) +\sum_{k,n,m} (\tilde{c}_{k,n,m} +r\tilde{q}_{k,n,m})  \\
\times (c_{k,n,m} +rq_{k,n,m}) + r (\tilde{c}_{k,n,m} - \tilde{q}_{k,n,m})(c_{k,n,m} -q_{k,n,m})
\end{array}
\end{sequation}%
We define $\boldsymbol{\Xi}'  = \mathbf{T}'(0)$, $ \boldsymbol{\Xi}  = \mathbf{T}'(-1)$,  $\mathbf{T}'(\tau) = {\rm{blockdiag}} \left(\mathbf{T}'_1(\tau), \mathbf{T}'_2(\tau), \dots,\mathbf{T}'_K(\tau)\right)$, and $\mathbf{T}'_k(\tau) = \sum_{k,m} \left(\sum_{n} g_{k,n,m}(\tau\tilde{c}_{k,n,m} + \tilde{q}_{k,n,m}) \right) \mathbf{T}_{k,m}$.
The parameters $\{c_{k,n,m},q_{k,n,m},\tilde{c}_{k,n,m},\tilde{q}_{k,n,m}\}$ are determined by equating the partial derivatives of ${\cal F}$ to zero. It is easy
to check that $\tilde{c}_{k,n,m} = 0, ~\forall k, n,m$ and $c_{k,n,m} = \tr (\mathbf{T}_{k,m}), ~\forall k,n,m$.

Motivated by the first term on the right side of (\ref{eq:TreplicaSym}) in the exponent, we can define a  Gaussian channel vector  as in (\ref{eq:EqScalGAUEach}). The conditional
distribution of the  Gaussian channel vector is given by (\ref{eq:hatx_k}). Upon the observation of the output $\mathbf{z}$, the optimal estimate of $\mathbf{x}$ in the mean-square
sense is
\begin{sequation}
\hat{\mathbf{x}}=E_{\mathbf{x}}\left[\mathbf{x}\left| \mathbf{z},\sqrt{{\boldsymbol{\Xi}}}\right.\right].
\end{sequation}%

Let $\gamma_{k,n,m} = \tilde{q}_{k,n,m}$ and $\psi_{k,n,m} = c_{k,n,m}-q_{k,n,m}$. Finally, at $r=0$, ${\cal F}$ can be expressed as
\begin{sequation}\label{eq:ap_GenFree}
{\cal F} \! = \! \ln 2 \ I\left( \mathbf{x};\mathbf{z} \big| \sqrt{\boldsymbol \Xi} \right)\! + \!\ln\det\left(\mathbf{I}_{N_r}+\mathbf{R} \right)\!- \!\sum_{k,n,m} \gamma_{k,n,m} \psi_{k,n,m} \! + \! N_r
\end{sequation}%
where $\boldsymbol{\Xi} = \mathbf{T}$, $\mathbf{T} =  {\rm{blockdiag}} \left(\mathbf{T}_1,\mathbf{T}_2, \dots,\mathbf{T}_K\right)$, $\mathbf{T}_k = \sum_{m}
\left(\sum_{n} g_{k,n,m} \gamma_{k,n,m} \right)\mathbf{T}_{k,m}$ and $\mathbf{R} = \sum_{k,n} \left(\sum_{m} \psi_{k,n,m}\right)$ $\mathbf{R}_{k,n}$.
The parameters $\gamma_{k,n,m}$ and $\psi_{k,n,m}$ are determined by equating the partial derivatives of ${\cal F}$ to zeros. Hence, we have
\begin{sequation} \label{eq:ap_gamma}
\gamma_{k,n,m} = \tr{\Big( \left(\mathbf{I}_{N_r}+\mathbf{R}\right)^{-1} \mathbf{R}_{k,n} \Big)}
\end{sequation}%
and
\begin{sequation}\label{eq:ap_fai}
\psi_{k,n,m} = \frac{\partial}{\partial \gamma_{k,n,m}}
 I\left(\mathbf{x};\mathbf{z} \left|\sqrt{\boldsymbol \Xi}\right.\right)
 = g_{k,n,m} \tr \left( \boldsymbol{\Omega}_k \mathbf{T}_{k,m} \right)
\end{sequation}%
where the derivative of the mutual information follows from the relationship between the mutual information and the MMSE revealed in \cite{Guo2005TIT_2,Palomar2006TIT}. Let $\gamma_{k,n} = \gamma_{k,n,m}$ and $\psi_{k,m} = \tr \left( \boldsymbol{\Omega}_k \mathbf{T}_{k,m} \right)$, for $m=1,2,\dots,M$.
Using (\ref{eq:ap_GenFree}) and substituting the
definitions of $\gamma_{k,n}$ and $\psi_{k,m}$, we then obtain (\ref{eq:GAUMutuall}) for the case $K_1 = K$.
The case with arbitrary value $K_1$ can be proved following a similar approach as above.



\end{document}